\documentclass[12pt,oneside,leqno]{article}

\usepackage{graphicx}
\usepackage[a4paper, margin=2cm]{geometry}

\usepackage[font=small,labelfont=bf,labelsep=period]{caption}

\usepackage[numbers,square,sort&compress]{natbib}

\usepackage{commath}
\usepackage{bm}

\newcommand{\nablab}{\bm{\nabla}}

\newcommand{\ub}{\bm{u}}
\newcommand{\eb}{\bm{e}}
\newcommand{\Mb}{\bm{M}}

\newcommand\T{\rule{0pt}{2.6ex}}       
\newcommand\B{\rule[-1.2ex]{0pt}{0pt}} 

\renewcommand{\eqref}[1]{Eq.~(\ref{#1})}
\renewcommand{\secref}[1]{Section~\ref{#1}}
\newcommand{\tabref}[1]{Table~\ref{#1}}

\makeatletter
\def\maketitle{
  {
   \fontsize{17}{20}\selectfont\sffamily{}  \noindent \MakeUppercase{\textbf{\@title}}

   \bigskip
   \fontsize{14}{20}\selectfont\rmfamily{} \noindent \@author
  }
}
\makeatother

\title{Geometric algebra and an acoustic space time for propagation in
  non-uniform flow}

\author{Alastair Gregory, Anurag Agarwal and Joan Lasenby
  \\
  {\small \textit{Department of Engineering, University of Cambridge,
    Trumpington Street, Cambridge, CB2 1PZ
    \\ email: alg57@cam.ac.uk}}
  \medskip
  \\
  Samuel Sinayoko
  \\
  {\small \textit{Institute of Sound and Vibration Research,
    University of Southampton, Highfield, Southampton, SO17 1BJ}}
}

\begin{document}

\maketitle

\renewcommand{\abstractname}{\vspace{-\baselineskip}} 

\begin{abstract}	\noindent
This study aims to make use of two concepts in the field of
aeroacoustics; an analogy with relativity, and Geometric Algebra. The
analogy with relativity has been investigated in physics and
cosmology, but less has been done to use this work in the field of
aeroacoustics. Despite being successfully applied to a variety of
fields, Geometric Algebra has yet to be applied to acoustics. Our aim
is to apply these concepts first to a simple problem in aeroacoustics,
sound propagation in uniform flow, and the more general problem of
acoustic propagation in non-uniform flows. By using Geometric Algebra
we are able to provide a simple geometric interpretation to a
transformation commonly used to solve for sound fields in uniform
flow. We are then able to extend this concept to an acoustic
space-time applicable to irrotational, barotropic background flows.
This geometrical framework is used to naturally derive the
requirements that must be satisfied by the background flow in order
for us to be able to solve for sound propagation in the non-uniform
flow using the simple wave equation. We show that this is not possible
in the most general situation, and provide an explicit expression that
must be satisfied for the transformation to exist. We show that this
requirement is automatically satisfied if the background flow is
incompressible or uniform, and for both these cases derive an explicit
transformation. In addition to a new physical interpretation for the
transformation, we show that unlike previous investigations, our work
is applicable to any frequency.
\end{abstract}

\section{The Acoustic Space-Time}
It was noted by Unruh \cite{Unruh:1981ue} that there is an analogy
between the equations of general relativity and acoustics in the
presence of background flow. This analogy has since been expanded on
by Visser \cite{Visser:1998wd} and Barcel{\'o} et al
\cite{Barcelo:2004vg} providing some illuminating space-time diagrams
of acoustic propagation in idealised flows. At the heart of this
analogy is the geometry within which the problem can be considered,
rather than a direct correspondence between governing equations.
Einstein's equations of gravitation (or the Einstein field equations)
\cite{Einstein:1920tx} are not the same as those that govern acoustic
perturbations (which we shall come to in a moment), however, the
geometry of the problem is usually considered to be a 4-dimensional
space of mixed signature which can in general be curved. It is
possible to consider acoustic perturbations in terms of a similar
space, and it is in this sense that there is an analogy.

If we consider an acoustic perturbation to a background flow,
described in terms of a perturbation velocity potential $\phi$, and
the background flow is irrotational and barotropic, then $\phi$ is
governed by
\cite{Taylor:1978dx,Unruh:1981ue,Visser:1998wd,PerezBergliaffa:2004bz},
\begin{equation}
  \label{eq:AST:irroBaroWaveEq}
  \del{\frac{\partial}{\partial t} + \ub\cdot\nablab}\frac{1}{c^2}
  \del{\frac{\partial}{\partial t} + \ub\cdot\nablab}\phi -
  \frac{1}{\rho}\nablab\cdot\del{\rho\nablab\phi} = 0,
\end{equation}
where $\ub$, $\rho$ and $c$ are the background flow velocity, density
and speed of sound, and $t$ and $\nablab$ are the standard time and
gradient operator of Newtonian mechanics. Conservation of mass for the
background flow tells us that,
\begin{equation}
  \frac{\partial\rho}{\partial t} + \nablab\cdot(\rho\ub) = 0.
\end{equation}
Using this along with \eqref{eq:AST:irroBaroWaveEq} we obtain a
slightly altered expression for $\phi$,
\begin{equation}
  \del{\frac{\partial}{\partial t} + \nablab\cdot\ub}\frac{\rho}{c^2}
  \del{\frac{\partial}{\partial t} + \ub\cdot\nablab}\phi -
  \nablab\cdot\del{\rho\nablab\phi} = 0,
\end{equation}
where $\nablab$ acts on \emph{everything} to its right. This may seem
like a needless complication , but as pointed out by, for example,
Visser \cite{Visser:1998wd}, we can write this as the d'Alembertian of
a curved, 4-dimensional space-time of mixed signature,
\begin{equation}
  \label{eq:AST:dAlembertian}
  \Delta\phi = \frac{1}{\sqrt{-g}}\partial_\mu
  \del{\sqrt{-g}g^{\mu\nu}\partial_\nu\phi} = 0.
\end{equation}
$g^{\mu\nu}$ is the inverse of the metric $g_{\mu\nu}$ of the space,
and $g$ is the determinant of $g_{\mu\nu}$. Hence it becomes clear
that we can regard $\phi$ as a scalar field on a 4-dimensional space
with the metric $g_{\mu\nu}$ defined by the background flow. This
space is the \emph{acoustic space-time}. The authors who have pointed
this out have largely used it as a tool to illuminate what parts of
general relativity are due to Einstein's field equations specifically,
and what phenomena remain when the equations are changed but the form
of the geometry remains (this is where black holes are of interest).
We, however, would like to turn this on its head, making use of the
acoustic space-time to illuminate how variable transformations can be
used to understand sound propagation in the presence of background
flow.

The idea of these transformations is to use a change of variables to
convert a challenging equation, such as \eqref{eq:AST:irroBaroWaveEq},
into, say, the classic wave equation.  To the authors knowledge the
most general transformation to date of this kind was first presented
by Taylor \cite{Taylor:1978dx}, and is valid for irrotational,
barotropic, low Mach number, steady background flows for acoustic
fields where the wavelength is of the same order of magnitude or
smaller than the length scale of variations of the background flow
\cite{Astley:1986ct}. The presentation of this transformation is
fairly ad hoc, and we would like to use the acoustic space-time to
derive and generalise such transformations in a more systematic way.

\section{Calculus on Curved Manifolds}
In order to proceed further we shall introduce some new tools to deal
with calculus on general manifolds. More precisely we shall introduce
Geometric Algebra (GA).

GA deals with general manifolds through the concept of embedding. A
flat vector space of large dimension is first defined, within which
the, possibly curved, manifold of interest is placed. The embedded
manifold then inherits a metric from the extrinsic space, and this
allows a study of Riemannian geometry.  Some readers may wonder why we
use GA instead of the more widely used differential forms and
differential geometry (see for example Nakahara
\cite{Nakahara:1990vx}); ultimately this a preference of the authors.
We find that the approach gives more streamlined and intuitive proofs
of some key results, and also allows more to be expressed
independently of coordinates. A debate of the relative benefits is
beyond the scope of the current paper. As was proven by John Nash in
1956, any finite dimensional Riemannian manifold can be embedded in a
larger dimensional flat space in such a way that the metric is
generated by the embedding \cite{Nash:1956vz}.  Therefore our approach
is easily adequate for our purposes, and all the results used should
still be familiar to anyone with knowledge of differential geometry.
For a full introduction to geometric calculus we direct the reader to
Hestenes \cite{Hestenes:1984vg} and Doran \cite{Doran:2003jd}. Key to
GA is the geometric product between vectors $ab=a\cdot b+a\wedge b$.
If two vectors appear adjacent then the geometric product is implied.

We consider an $n$ dimensional curved manifold embedded within a flat
space of higher dimension. Let $x^i$ be a set of $n$ scalar coordinate
functions defined over the part of the manifold of interest and $x$ be
some point on the manifold. A basis of the tangent space (frame) of
the manifold is given by $e_i=\partial x/\partial x^i$. The reciprocal
frame $\{e^i\}$ of the tangent space is defined to satisfy,
\begin{equation}
  \label{eq:CCM:recipReq}
  e_j\cdot e^i = \delta^i_j
\end{equation}
where the inner product is inherited from the flat embedding space.
The vector derivative intrinsic to the manifold $\partial$ is defined
as the projection of the vector derivative of the embedding space onto
the manifold, and can be written as,
\begin{equation}
  \label{eq:CCM:partialDefn}
  \partial = e^i\frac{\partial}{\partial x^i} =
  e^i\partial_{x^i}.
\end{equation}
In addition we can show that the reciprocal frame $\{e^i\}$ is given
by,
\begin{equation}
  \label{eq:CCM:coordTrans}
  e^i = \partial x^i.
\end{equation}
Let the local pseudoscalar of the manifold be $I(x)$. We define the
scalar field $V(x)$, which is specific to the frame $\{e_i\}$, such
that it satisfies,
\begin{equation}
  e_1\wedge e_2\wedge\cdots\wedge e_n = VI.
\end{equation}
The covariant derivative $DA$ of a multivector field $A$ (defined on
the manifold) is defined as the projection onto the manifold of
$\partial A$. A useful result for a scalar field $\alpha$ is that,
\begin{equation}
  \label{eq:CCM:genLaplace}
  D\cdot(D\alpha) = D^2\alpha = \frac{1}{V}
  \frac{\partial}{\partial x^i}\del{Vg^{ij}
    \frac{\partial\alpha}{\partial x^j}},
\end{equation}
where $g^{ij}=e^i\cdot e^j$. Similarly $g_{ij}=e_i\cdot e_j$, and this
satisfies $g^{ik}g_{kj}=\delta^i_j$. We can also show that
$V=\sqrt{\abs{\det g_{ij}}}$. $g_{ij}$ is the metric of the frame.

\section{A Transformation Derived from the Acoustic Space-Time}
\subsection{A Preliminary Investigation into Uniform Flow}
\label{sec:UF}
To begin we consider a 4-dimensional flat acoustic space-time. We
define three frames within this which satisfy,
\begin{equation}
  \label{eq:UniFlow:Metrics}
  \begin{array}{c  c  c}
    \{e_i\} & \{e_i'\} & \{e_i''\} \\
    V = 1 & V' = 1 & V'' = 1 \\
    g^{ij} = \begin{bmatrix}1&0&0&M\\0&-1&0&0\\
      0&0&-1&0\\M&0&0&M^2-1\end{bmatrix} &
    {g^{ij}}' = \begin{bmatrix}1&0&0&0\\0&-1&0&0\\
      0&0&-1&0\\0&0&0&-1\end{bmatrix} &
    {g^{ij}}'' = \begin{bmatrix}1&0&0&0\\0&-1&0&0\\
      0&0&-1&0\\0&0&0&-1\end{bmatrix}
  \end{array}
\end{equation}
and are related to each other by,
\begin{subequations}
\begin{gather}
  \label{eq:UniFlow:GaliTransFrame}
  e_0 = e_0'' - Me_z'',\quad e_1 = e_1'',\quad
  e_2 = e_2'',\quad e_3 = e_3'', \\
  e_0' = \frac{1}{\beta}\del{e_0''-Me_3''},\quad
  e_1' = e_1'',\quad e_2' = e_2'',\quad
  e_3' = \frac{1}{\beta}\del{e_3''-Me_0''},
\end{gather}
\end{subequations}
where $M$ is a constant and $\beta=\sqrt{1-M^2}$ (it is simple to
check that these transformations are consistent with the metrics
defined above). From this we can derive the following coordinate
transformations,
\begin{subequations}
\begin{gather}
  \label{eq:UniFlow:GaliTrans}
  x^0 = {x^0}'',\quad x^1 = {x^1}'',\quad x^2 = {x^2}'',\quad
  x^3 = {x^3}'' + M{x^0}'', \\
  \label{eq:UniFlow:LorTrans}
  {x^0}' = \frac{1}{\beta}\del{{x^0}'' + M{x^3}''},\quad
  {x^1}' = {x^1}'',\quad {x^2}' = {x^2}'',\quad
  {x^3}' = \frac{1}{\beta}\del{{x^3}'' + M{x^0}''},
\end{gather}
\end{subequations}

If $c_0$ is a constant, and we define the following coordinates of the
three frames,
\begin{equation}
  x^0=c_0t,\quad x^1=x,\quad x^2=y,\quad x^3=z,
\end{equation}
with similar relations for $\{{x^i}'\}$ and $\{{x^i}''\}$. From
\eqref{eq:CCM:genLaplace} $D^2\phi=0$ may be written in the 3 frames
as,
\begin{subequations}
\begin{align}
  \label{eq:UniFlow:ObsFrWaveEq}
  0 = D^2\phi &= \sbr{\tfrac{1}{c_0^2}\del{
    \partial_t + U\partial_z}^2 -
  \partial_x^2 - \partial_y^2 - \partial_z^2}\phi \\
  \label{eq:UniFlow:LorFrWaveEq}
  &= \sbr{\tfrac{1}{c_0^2}\partial_{t'}^2 -
  \partial_{x'}^2 - \partial_{y'}^2 - \partial_{z'}^2}\phi \\
  \label{eq:UniFlow:FluidFrWaveEq}
  &= \sbr{\tfrac{1}{c_0^2}\partial_{t''}^2 -
  \partial_{x''}^2 - \partial_{y''}^2 - \partial_{z''}^2}\phi,
\end{align}
\end{subequations}
where $U=Mc_0$. From this we see that \eqref{eq:UniFlow:ObsFrWaveEq}
is identical to \eqref{eq:AST:irroBaroWaveEq} when the background flow
is uniform, while \eqref{eq:UniFlow:FluidFrWaveEq} is identical to
\eqref{eq:AST:irroBaroWaveEq} when there is no flow. This result is
expected since the transformation from $\{e_i\}$ to $\{e_i''\}$ is a
Galilean transformation. $\{e_i\}$ represents a frame moving with Mach
number $M$ relative to the fluid in the $-z$ direction, while
$\{e_i''\}$ represents the fluid frame.

Inspecting \eqref{eq:UniFlow:LorTrans} we see that the transformation
from $\{e_i''\}$ to $\{e_i'\}$ is a Lorentz transform, and so it is
not surprising that the metric is unchanged, and from
\eqref{eq:CCM:genLaplace} we see that $D^2\phi$ will take the simplest
form. However, by deriving the relations between the $x^i$ and
${x^i}'$ coordinates,
\begin{equation}
  \label{eq:UniFlow:FullTrans}
  {x^0}' = \beta x^0 + \frac{M}{\beta}x^3,\quad
  {x^1}' = x^1,\quad {x^2}' = x^2,\quad
  {x^3}' = \frac{x^3}{\beta},
\end{equation}
we can see that if an observer is stationary in the $\{e_i\}$ frame,
they will also be stationary in the $\{e_i'\}$ frame. The value of the
$\{e_i'\}$ frame lies in the fact that it has a simple metric, and so
a simple form of the wave equation, but moves with the observer frame.
The transformation in \eqref{eq:UniFlow:FullTrans} has been presented
before by, for example, Chapman \cite{Chapman:2000ky}, but by
interpreting it though an acoustic space time we are able to see why
the transformation works, and precisely what frame we are transforming
to when we use it.

\subsection{Motivation: Generalised Galilean and Lorentz Transforms}
Now we would like to try to generalise this process to non-uniform
flows. We start from the frame defined in \secref{sec:UF},
$\{e_i''\}$. Note that for now we are still dealing with a flat
acoustic space-time. We would like to now find a frame in which
\eqref{eq:CCM:genLaplace} gives \eqref{eq:AST:irroBaroWaveEq} in the
general case.

\subsubsection{Generalised Galilean Transform}
\label{sec:NUF:GalileanTrans}
We redefine the $\{e_i\}$ frame and its reciprocal as,
\begin{equation}
  \label{eq:NUF:observerFrame}
  \begin{gathered}
    e_0 = h_\tau\del{e_0''-M_1e_1''-M_2e_2''-M_3e_3''},\quad
    e_1 = he_1'',\quad e_2 = he_2'',\quad e_3 = he_3'',
  \end{gathered}
\end{equation}
where $h_\tau=\sqrt{\rho c/\rho_0c_0}$ and $h=\sqrt{\rho c_0/\rho_0
  c}$. We interpret the $M_i$ as Mach number components, $\rho$ and
$c$ as the (non-uniform) background density and speed of sound, and
$\rho_0$ and $c_0$ as the (constant) free-field density and speed of
sound.  Comparing this to \eqref{eq:UniFlow:GaliTransFrame} we see
that this is a generalised form of a Galilean transform. For this
frame we can show that $V=h^3h_\tau$ and,
\begin{equation}
  Vg^{ij} = \frac{\rho}{\rho_0c^2}\begin{bmatrix}
    c_0^2 & c_0u_1 & c_0u_2 & c_0u_3 \\
    c_0u_1 & u_1^2-c^2 & u_1u_2 & u_1u_3 \\
    c_0u_2 & u_1u_2 & u_2^2-c^2 & u_2u_3 \\
    c_0u_3 & u_1u_3 & u_2u_3 & u_3^2-c^2
  \end{bmatrix},
\end{equation}
where $u_i=M_ic$ are interpreted as the velocity components of $\ub$
in \eqref{eq:AST:irroBaroWaveEq}. If we denote the coordinates
associated with the $\{e_i\}$ frame as $x^0=c_0t,x^1=x,x^2=y,x^3=z$,
then by \eqref{eq:CCM:genLaplace} $D^2\phi=0$ may be written as,
\begin{equation}
  \label{eq:NUF:observerFrameWaveEq}
  \begin{aligned}
    D^2\phi &= \frac{1}{V\rho_0}\left[\partial_t(\rho/c^2)\del{
      \partial_t+u_1\partial_x+u_2\partial_y+u_3\partial_z}\right.\\
      &\qquad + \partial_x(\rho/c^2)\del{u_1\partial_t +
      (u_1^2-c^2)\partial_x + u_1u_2\partial_y + u_1u_3\partial_z} \\
      &\qquad + \partial_y(\rho/c^2)\del{u_2\partial_t +
      u_1u_2\partial_x + (u_2^2-c^2)\partial_y + u_2u_3\partial_z} \\
      &\qquad + \left.\partial_z(\rho/c^2)\del{u_3\partial_t +
      u_1u_3\partial_x + u_2u_3\partial_y + (u_3^2-c^2)\partial_z}
      \right] \phi = 0,
  \end{aligned}
\end{equation}
which is a multiple of \eqref{eq:AST:irroBaroWaveEq}, as required.

\subsubsection{Generalised Lorentz Transform}
Following the methodology of \secref{sec:UF}, we will now try to produce a
frame that moves with $\{e_i\}$, but that has a simple metric, and so
simplifies the governing equation. Here the notation of geometric
algebra will be very useful. To produce a frame with a simple metric
we will apply a Lorentz transform to the $\{e_i''\}$ frame, since a
Lorentz transform preserves the metric. Furthermore we shall apply a
Lorentz transform that produces a frame whose time vector ($e_0'$) is
parallel to $e_0$ from the previous section. It is in this sense that
the $\{e_i'\}$ frame will move with the $\{e_i\}$ frame.

In geometric algebra, Lorentz transforms have a very neat
representation in the form of rotors (see \cite[\S5.4]{Doran:2003jd}).
Hence we can produce the frame just described using the simple
relation,
\begin{equation}
  e_i' = R e_i'' \tilde R,
\end{equation}
where $R$ is the rotor and $\tilde R$ denotes the reverse of $R$ (see
\cite[\S4.1.3]{Doran:2003jd}). $R$ is given explicitly by the
expression,
\begin{equation}
  \begin{gathered}
    R = \exp\del{-\frac{\alpha}{2}e_me_0''}, \\
    \tanh\alpha=M,\quad
    e_m=(M_1e_1''+M_2e_2''+M_3e_3'')/M,\quad
    M^2 = M_1^2+M_2^2+M_3^2.
  \end{gathered}
\end{equation}
Applying this transformation we obtain the $\{e_i'\}$ frame,
\begin{equation}
  \label{eq:NUF:LorentzFrame}
  \begin{gathered}
    e_0' = \frac{1}{\beta}\del{e_0''-Me_m},\quad
    e_1' = -\frac{M_1}{\beta}e_0'' + e_1 +
    \frac{M_1}{M}\frac{1-\beta}{\beta}e_m, \\
    e_2' = -\frac{M_2}{\beta}e_0'' + e_2 +
    \frac{M_2}{M}\frac{1-\beta}{\beta}e_m,\quad
    e_3' = -\frac{M_3}{\beta}e_0'' + e_3 +
    \frac{M_3}{M}\frac{1-\beta}{\beta}e_m.
  \end{gathered}
\end{equation}
Comparing these definitions with \eqref{eq:NUF:observerFrame} we see
that $e_0'$ is indeed parallel to $e_0$. Furthermore it is relatively
simple to show that the reciprocal frame $\{{e^i}'\}$, volume form $V'$
and metric ${g^{ij}}'$ are the same as in \eqref{eq:UniFlow:Metrics}.
From this it follows that $D^2\phi=0$ can be written as
\eqref{eq:UniFlow:LorFrWaveEq} in this frame.

So far we have been working in a flat acoustic space-time. Despite
this, we appear to have found a frame where we can write
\eqref{eq:AST:irroBaroWaveEq} independently of the frame. However, we
have so far skipped over deriving the explicit coordinate
transformations from (say) $\{{x^i}''\}$ to $\{x^i\}$. To do this, we
must use \eqref{eq:CCM:coordTrans}. In fact, by assuming a flat
space-time, we have made it possible for the coordinate transformation
to not exist.  To get around this problem, we will now generalise to a
curved space time. The reader may wonder why we went to the trouble of
deriving the results for a flat space time, but we will see in the
next section that the frame transformations in
\eqref{eq:NUF:observerFrame} and \eqref{eq:NUF:LorentzFrame} will in
fact be very useful, and without taking our detour through the flat
space-time, there is little chance that we would have arrived at these
results.

\subsection{Transforming in a Curved Acoustic Space-Time}
We consider a 4-dimensional (possibly curved) manifold with an
associated coordinate system $\{x^i\}$ and frame $\{e_i\}$ defined
such that the metric $g_{ij}$ is given by,
\begin{equation}
  e_i\cdot e_j = g_{ij} = \begin{bmatrix}
    h_\tau^2\beta^2 & M_1h_\tau h & M_2h_\tau h& M_3h_\tau h \\
    M_1h_\tau h & -h^2 & 0 & 0 \\
    M_2h_\tau h & 0 & -h^2 & 0 \\
    M_3h_\tau h & 0 & 0 & -h^2
  \end{bmatrix},
\end{equation}
where symbols have been defined in \secref{sec:NUF:GalileanTrans}. We
can show that the inverse metric $g^{ij}$ and the volume form $V$ are
given in this case by,
\begin{equation}
  V = h^3h_\tau = \frac{\rho^2c_0}{\rho_0^2c},\quad
  g^{ij} = \frac{\rho_0}{\rho cc_0}\begin{bmatrix}
    c_0^2 & c_0u_1 & c_0u_2 & c_0u_3 \\
    c_0u_1 & u_1^2-c^2 & u_1u_2 & u_1u_3 \\
    c_0u_2 & u_1u_2 & u_2^2-c^2 & u_2u_3 \\
    c_0u_3 & u_1u_3 & u_2u_3 & u_3^2-c^2
  \end{bmatrix}
\end{equation}
From this it is simple to show that $D^2\phi=0$ written out in this
frame is \eqref{eq:NUF:observerFrameWaveEq} which we have already
noted is a multiple of \eqref{eq:AST:irroBaroWaveEq} if we define the
coordinates $x^0=c_0t,x^1=x,x^2=y,x^3=z$.

We now need to define the frame $\{e_i'\}$ that spans the tangent
space spanned by $\{e_i\}$ and has a simple metric. Now the reader
should see why we went to all the trouble of generalising the Galilean
and Lorentz transforms. Using \eqref{eq:NUF:observerFrame} and
\eqref{eq:NUF:LorentzFrame} we can find precisely this frame
transformation, which we can still use locally in the tangent space of
our newly defined curved manifold. Hence we define the frame
$\{e_i'\}$ as,
\begin{equation}
  \begin{gathered}
    e_0' = \frac{1}{h_\tau\beta}e_0,\quad
    e_1' = \frac{1}{h}e_1 +
    \frac{M_1}{\beta h}\del{\frac{1-\beta}{M^2}-1}m -
    \frac{M_1}{\beta h_\tau}e_0, \\
    e_2' = \frac{1}{h}e_2 +
    \frac{M_2}{\beta h}\del{\frac{1-\beta}{M^2}-1}m -
    \frac{M_2}{\beta h_\tau}e_0,\quad
    e_3' = \frac{1}{h}e_3 +
    \frac{M_3}{\beta h}\del{\frac{1-\beta}{M^2}-1}m -
    \frac{M_3}{\beta h_\tau}e_0,
  \end{gathered}
\end{equation}
where $m$ is defined as $m = M_1e_1 + M_2e_2 + M_3e_3$. We can show
that the volume measure $V'$ and metric ${g^{ij}}'$ for this frame are
the same as those given in \eqref{eq:UniFlow:Metrics}, as expected.
$D^2\phi=0$ written in this frame gives
\eqref{eq:UniFlow:LorFrWaveEq}.

Now we relate the coordinates of the $\{e_i\}$ frame to those of the
$\{e_i'\}$ frame. From
Eqs.~(\ref{eq:CCM:recipReq}),~(\ref{eq:CCM:partialDefn})~and~(\ref{eq:CCM:coordTrans}),
\begin{equation}
  \label{eq:NUF:coordTrans}
  {e^i}' = \partial({x^i}') = e^k\partial_{x^k}({x^i}')
  \;\Rightarrow\;
  \frac{\partial{x^i}'}{\partial x^j} = {e^i}'\cdot e_j.
\end{equation}
Therefore, using the frame definitions given in this section, we can
derive the set of relationships given in \tabref{tab:NUF:coordRel}.
The coordinate transformation from $\{x^i\}$ to $\{{x^i}'\}$ must
satisfy these relationships.

\begin{table}
  \centering
  \caption{Relationships between the $\{x^i\}$ and $\{{x^i}'\}$
    coordinates, found using \eqref{eq:NUF:coordTrans}.}
  \begin{tabular}{c c | c c}
    \hline\hline
    $\partial{x^0}'/\partial x^0$ & $h_\tau\beta$ &
    $\partial{x^0}'/\partial x^1$ & $hM_1/\beta$
    \T\\
    $\partial{x^1}'/\partial x^0$ & $0$ &
    $\partial{x^1}'/\partial x^1$ &
    $h(M_1^2+\beta M_2^2+\beta M_3^2)/(M^2\beta)$
    \\
    $\partial{x^2}'/\partial x^0$ & $0$ &
    $\partial{x^2}'/\partial x^1$ &
    $h[M_1M_2(1-\beta)]/(M^2\beta)$
    \\
    $\partial{x^3}'/\partial x^0$ & $0$ &
    $\partial{x^3}'/\partial x^1$ &
    $h[M_1M_3(1-\beta)]/(M^2\beta)$
    \B\\\hline
    $\partial{x^0}'/\partial x^2$ & $hM_2/\beta$ &
    $\partial{x^0}'/\partial x^3$ & $hM_3/\beta$
    \T\\
    $\partial{x^1}'/\partial x^2$ &
    $h[M_1M_2(1-\beta)]/(M^2\beta)$ &
    $\partial{x^1}'/\partial x^3$ &
    $h[M_1M_3(1-\beta)]/(M^2\beta)$
    \\
    $\partial{x^2}'/\partial x^2$ &
    $h(\beta M_1^2+M_2^2+\beta M_3^2)/(M^2\beta)$ &
    $\partial{x^2}'/\partial x^3$ &
    $h[M_2M_3(1-\beta)]/(M^2\beta)$ \\
    $\partial{x^3}'/\partial x^2$ &
    $h[M_2M_3(1-\beta)]/(M^2\beta)$ &
    $\partial{x^3}'/\partial x^3$ &
    $h(\beta M_1^2+\beta M_2^2+M_3^2)/(M^2\beta)$
    \B\\
    \hline\hline
  \end{tabular}
  \label{tab:NUF:coordRel}
\end{table}

\section{Further Manipulations of the Transform}
We have derived the 16 partial differential equations that a
transformation must satisfy (see \tabref{tab:NUF:coordRel}) in terms
of the observable coordinates $\{x^i\}$ and the transformed
coordinates $\{{x^i}'\}$. These are all scalar fields, and we now map
these to a 3-dimensional Euclidean space plus time (the standard
manifold of Newtonian mechanics). This allows us to write the
relations given in \tabref{tab:NUF:coordRel} in a more illuminating
way.

We treat $x^1,x^2,x^3$ as Cartesian coordinates in the three
dimensional space, $t=x^0/c_0$ as time, and the transformed
coordinates $\{{x^i}'\}$ as scalar fields over this space (note that
this map between the acoustic space-time and a Euclidean space is
precisely the map needed to take us from the equation $D^2\phi=0$ to
\eqref{eq:AST:irroBaroWaveEq}). We denote the Cartesian basis vectors
of this space $\{\eb_1,\eb_2,\eb_3\}$. If we denote the vector
derivative of the 3-dimensional space as $\nablab$, and define the
Mach number vector $\Mb$ as $\Mb=M_i\eb_i$, then we can show that the
relations in \tabref{tab:NUF:coordRel} can be written,
\begin{equation}
  \label{eq:FMT:coorTransEuclid}
  \begin{gathered}
    \frac{\partial{x^0}'}{\partial x^0} = h_\tau\beta,\quad
    \frac{\partial{x^1}'}{\partial x^0} = 0,\quad
    \nablab{x^0}' = \frac{h}{\beta}\Mb,\quad
    \nablab{x^1}' = h\frac{1-\beta}{M^2\beta}M_1\Mb + h\eb_1, \\
    \frac{\partial{x^2}'}{\partial x^0} = 0,\quad
    \frac{\partial{x^3}'}{\partial x^0} = 0,\quad
    \nablab{x^2}' = h\frac{1-\beta}{M^2\beta}M_2\Mb + h\eb_2,\quad
    \nablab{x^3}' = h\frac{1-\beta}{M^2\beta}M_3\Mb + h\eb_3,
  \end{gathered}
\end{equation}
Combining the expressions for $\nablab{x^0}'$, $\nablab{x^1}'$,
$\partial{x^0}'/\partial x^0$ and $\partial{x^1}'/\partial x^0$ we
can derive,
\begin{equation}
  \frac{\partial}{\partial x^0}\del{\frac{1-\beta}{M^2}M_1}
  \frac{h}{\beta}\Mb + \frac{1-\beta}{M^2}M_1\nablab(h_\tau\beta) +
  \frac{\partial h}{\partial x^0}\eb_1 = 0.
\end{equation}
This is a constraint that the background flow must satisfy if the
transformation is to exist. We have not been able to show that this is
satisfied in the general case, for instance we cannot find any
equivalence between this and the compressible potential flow
equations. This would seem to suggest that a transformation is not
possible in the general case, however, there are some special cases
still of interest. If we assume that the background flow is steady
then the condition that must be satisfied in order for the
transformation to exist becomes,
\begin{equation}
  \nablab(h_\tau\beta)=0 \quad \Rightarrow \quad
  \nablab\del{\sqrt{\rho c(1-M^2)}} = 0.
\end{equation}
Again, we have been unable to show any equivalence between this and
the compressible potential flow equations, however it is clear that,
if the flow is uniform, or incompressible (low Mach number), then this
will be satisfied. In both of these cases $h=h_\tau=1$. For uniform
flow the relations in \tabref{tab:NUF:coordRel} can be integrated
directly. If we also assume that flow is only in the $z$ direction
then we obtain the transformation given in
\eqref{eq:UniFlow:FullTrans}, as expected. For incompressible steady
flow, the relations in \eqref{eq:FMT:coorTransEuclid} simplify to
become,
\begin{equation}
  \frac{\partial{x^0}'}{\partial x^0} = 1,\quad
  \frac{\partial{x^i}'}{\partial x^0} = 0,\quad
  \nablab{x^0}' = \Mb,\quad
  \nablab{x^i}' = \eb_i,
\end{equation}
where $i=1,2,3$. The background flow is potential, so we may write
$\Mb=\nablab\Phi$. From this we see that the transformation for an
incompressible, irrotational, steady, barotropic background flow is,
\begin{equation}
  \label{eq:FMT:TaylorTrans}
  {x^0}' = x^0 + \Phi,\quad {x^1}'=x^1,\quad
  {x^2}'=x^2,\quad {x^3}'=x^3.
\end{equation}
This agrees with a result presented by Taylor \cite{Taylor:1978dx},
however, as pointed out by Astley \cite{Astley:1986ct}, Taylor had to
make an assumption about the frequency of the acoustic field in order
to derive the transform. We have removed this requirement,
generalising Taylor's transform to all frequencies.

\section{Conclusions}
An acoustic space-time was used to derive the requirements that must be
satisfied by a variable transformation to solve for sound propagation
in the presence of irrotational, barotropic background flow. By
comparison to previous work deriving such transformations our approach
is more systematic, generalising the combined Galilean and Lorentz
transformations used when the flow is uniform.

We have shown that in order for the transformation to exist the
background flow must satisfy additional constraints to the
compressible potential flow equations, and that these additional
constraints are satisfied automatically if the background flow is
steady, and either uniform, or incompressible. In the former case we
have shown that the derived transformation agrees with the standard
transformation for uniform flow. In the latter case we show that our
transformation agrees with one presented before by Taylor; but, unlike
Taylor, we do not need to make any assumption about the frequency of
the acoustic field, and so we have extended the validity to low
frequency acoustic problems.

\renewcommand\refname{REFERENCES}
\bibliographystyle{rspub_unsrt}
\bibliography{GAbib}

\end{document}